\documentclass{ceab}   

\usepackage{epsfig}     
\usepackage{graphicx}   

\usepackage{ceabbib}     

\usepackage{amsmath}%
\usepackage{amssymb}

\begin{document}

\title{Spatial restriction to HXR footpoint locations by reconnection site geometries}

\author{M. TEMMER$^1$, B. VR\v{S}NAK$^1$, A. VERONIG$^{2}$, and M.
MIKLENIC$^{2,}$ \vspace{2mm}\\ \it $^1$Hvar Observatory, Faculty
of Geodesy, University of Zagreb,\\ \it Ka\v{c}i\'{c}eva 26,
HR--10000 Zagreb, Croatia\\ \it $^2$IGAM/Kanzelh\"ohe Observatory,
Institute of Physics, University of Graz\\ \it Universit\"atsplatz
5, A-8010 Graz, Austria}

\maketitle
\def\gore{Restriction to HXR footpoint locations}

\begin{abstract}
It is assumed that HXR sources map to the primary energy release
site in flares where particle acceleration occurs. Strong HXR
sources are mostly observed at confined regions along the
reconnecting magnetic arcade. We make a general approach on how
the geometry of the reconnecting current sheet (CS) may influence
the strength and localization of observed HXR sources. For this we
use results from an analysis on the 3B/X3.8 flare on January 17,
2005 \citep{temmer07}, as well as measurements from the associated
CME. Due to the close match of the CME acceleration profile and
the flare HXR flux, we suppose that the CME might play a certain
role in modifying the geometry of the CS (``symmetric'' versus
``asymmetric'' vertically stretched CS). This could be the driver
for ``guiding'' the accelerated particles to confined areas along
the flaring arcade and might explain the spatially limited
occurrence of strong HXR sources in comparison to elongated
ribbons as seen in H$\alpha$ and UV.

\end{abstract}


\section{Introduction}

It is generally accepted that hard X-ray (HXR) sources map to the
primary energy release site in solar flares, where particle
acceleration is assumed to occur \citep[e.g.][]{fletcher01}, and
thus give insight into the energy release process. The flare HXR
emission is mainly concentrated at the footpoints of magnetic
loops \cite[e.g.,][]{HoyngEA81,Sakao94}, and is assumed to be
produced by accelerated electrons that are collisionally stopped
in the `dense' chromosphere (as compared to the tenuous corona)
and emit nonthermal thick-target bremsstrahlung when braking in
the field of the ions \citep{Brown71}. For some events there is
also evidence for nonthermal hard X-rays from the corona
\citep{MasudaEA94,LinEA03,VeronigBrown04}.

When comparing HXR sources with H$\alpha$ and UV images it is
often observed that the HXR emission is concentrated in some
compact sources, which cover only a small part of the flare
ribbon, and are predominantly associated with bright H$\alpha$
(UV) kernels located on the outer edge of the ribbons
\citep[e.g.,][]{HoyngEA81,SakaoEA92,asai04,kasparova05,KruckerEA05}.
There are only rare exceptions that seem to show ``HXR ribbons"
\cite[for an example see][]{MasudaEA01}.

A case study by \cite{asai02,asai04} using Yohkoh HXR images,
showed that this discrepancy might be due to the limited dynamic
range of X-ray instruments. This may lead to the effect that only
the strongest nonthermal sources are observed in HXRs, whereas the
weaker ones are buried in the noise of the instruments. We
performed a similar study on the X3.8 flare of January 17, 2005
\citep[][hereafter referred to as Paper~I]{temmer07} where we
compared the locations of HXR sources as observed with the RHESSI
instrument to the occurrence of H$\alpha$ ribbons (using data from
the Hvar and Kanzelh\"ohe Observatory). From this we found that
the accelerated electrons are preferentially focused into a small
subset of loops (outlined by the HXR footpoints) of all the loops
that take part in the magnetic reconnection process (outlined by
the H$\alpha$ ribbons and EUV postflare arcade). Comparing the
local energy release and reconnection rates for H$\alpha$ flare
ribbon locations accompanied by HXR footpoints and those without,
we found differences of two and one order of magnitude,
respectively. These differences are large enough to explain the
different flare morphologies typically observed in HXRs (compact
footpoints) and H$\alpha$/UV (extended ribbons) by the limited
dynamic range of the RHESSI HXR instrument.

The following analysis is based on the results presented in Paper~I
from the X3.8 flare of January 17, 2005. In this paper we aim to
make a more general approach to the determination of specific HXR
locations. We will make use of measurements from the associated
coronal mass ejection and try to show how the geometry of the
reconnecting large scale current sheet might influence the
location of strong HXR sources.

\section{CME measurements and geometry of the current sheet}

A classical two ribbon flare is generally explained within the
two-dimensional ``CHSKP'' model \citep{cm64,st66,hir74,kopp76} in
which a prominence rises, stretches the field lines underneath,
and forms a current sheet (CS). Due to various instabilities a
sudden increase in the resistivity takes places after exceeding a
critical value of the CS length \citep[see e.g.][]{treumann97}.
Then the reconnection process starts in the region of enhanced
resistivity (\textit{diffusion region}). The coronal magnetic
field lines that successively reconnect result in a growing
flare/postflare loop system and separating H$\alpha$ and UV flare
ribbons, as observed in many flares \citep[e.g.,][]{SvestkaEA87,
TsunetaEA92, Svestka96, fletcher01, KruckerEA03, asai04, SuiEA04,
VeronigEA06, VrsnakEA06}.

\begin{figure}
  \begin{center}
   \epsfig{file=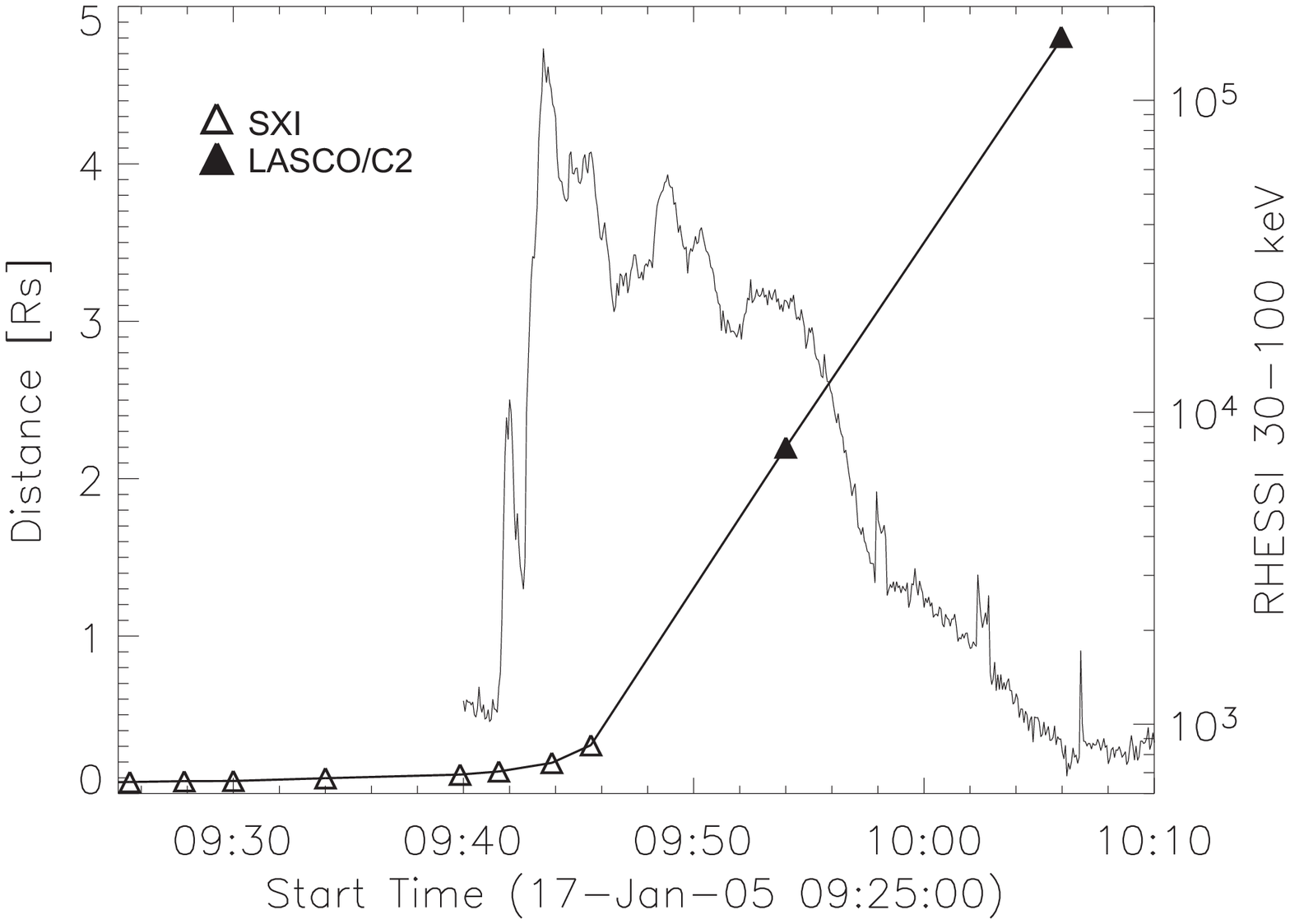,width=10cm}\\
  \epsfig{file=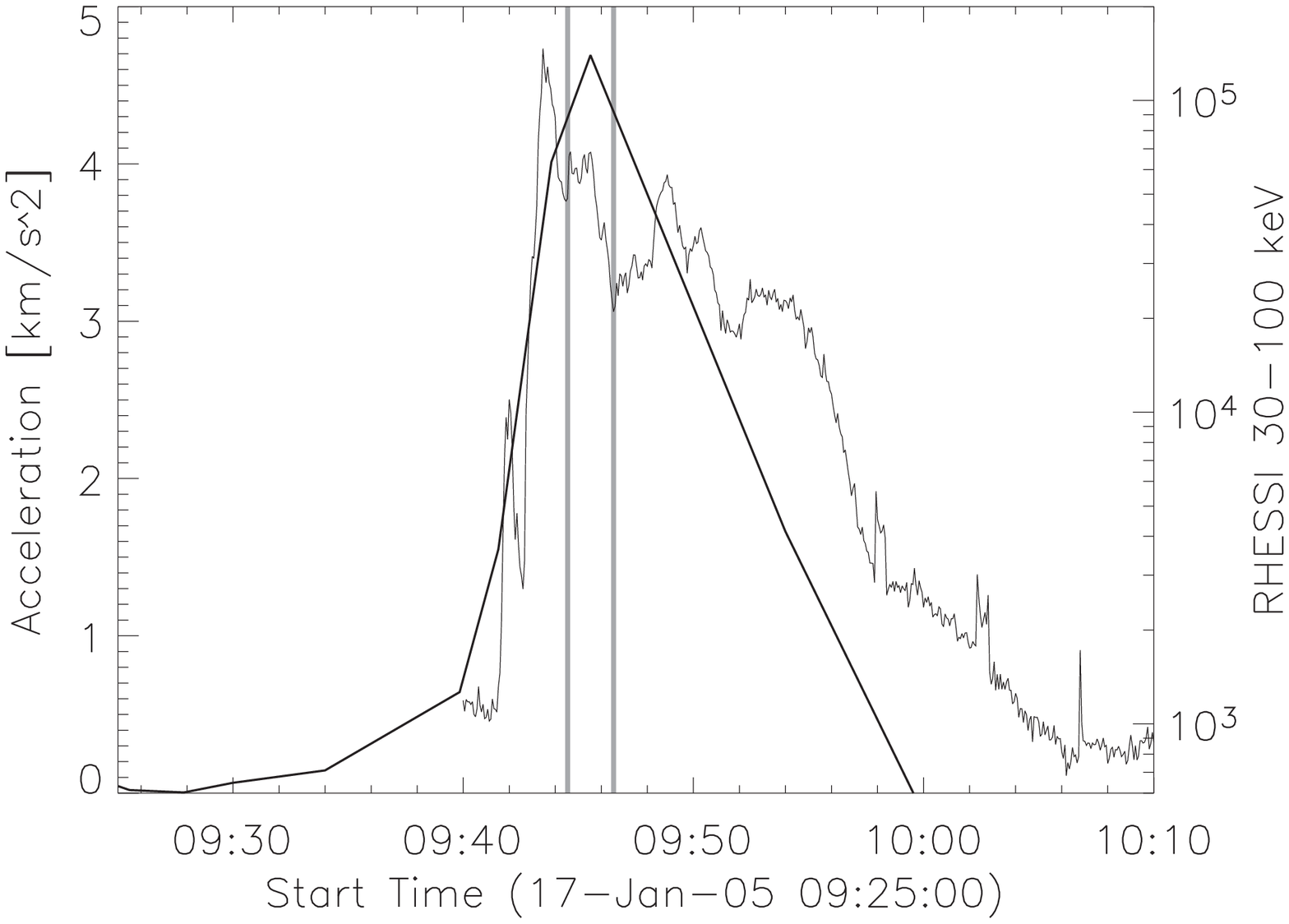,width=10cm}%
 \end{center}
  \caption{\textit{Left}: Height-time measurements of CME observations in comparison to the
HXR flux (RHESSI 30--100~keV energy range). \textit{Right}:
Acceleration of the CME in comparison to the HXR flux (RHESSI
30--100~keV energy range). CME measurements are from SXI and
LASCO/C2 observations.}\label{cme}
\end{figure}

The January 17, 2005 X3.8 flare was accompanied by a halo coronal
mass ejection (CME). From GOES/SXI observations we could follow
the rising of the flux rope which can be seen as a proxy for the
CME first phase in the low corona. As shown in the left panel of
Fig.~\ref{cme} we measured the CME from its initiation height at
$\sim$0.07 up to $\sim$0.3~$R_{\odot}$ from SXI and in the range
of $\sim$2--5~$R_{\odot}$ in two LASCO C2 images. The linear fit
to the height-time measurements from 9:42--10:06~UT gives a mean
speed of $\overline{v}\sim$1900~km/s. The right panel of
Fig.~\ref{cme} shows the acceleration of the CME derived from the
second derivative to the height-time measurements. As it can be
seen, the CME acceleration profile is very similar to the HXR flux
which gives evidence of a high relation between the flare energy
release and the CME acceleration\footnote{We want to note that the
peak time of the acceleration profile is highly dependent on the
measured CME features which are assumed to be the same in SXI and
LASCO observations. To get an error estimation for this we
measured the most conservative and most speculative features in
the LASCO observations for which we obtain $a_{max}$$\pm$1~min
indicated in Fig.~\ref{cme} by gray vertical lines.} \citep[see
also][]{vrsnak04}.

\begin{figure}
  \begin{center}
\epsfig{file=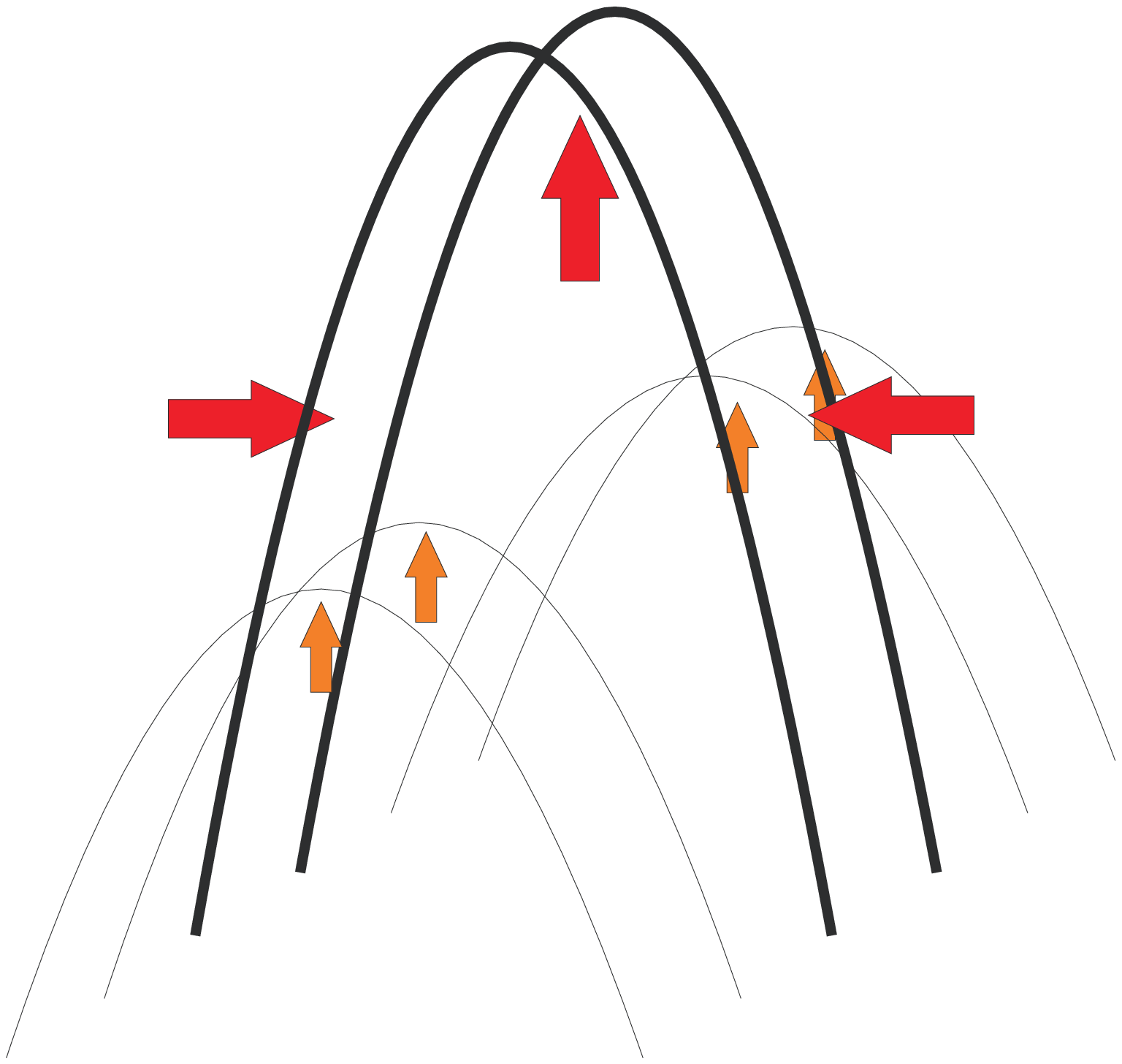,width=5cm}
\epsfig{file=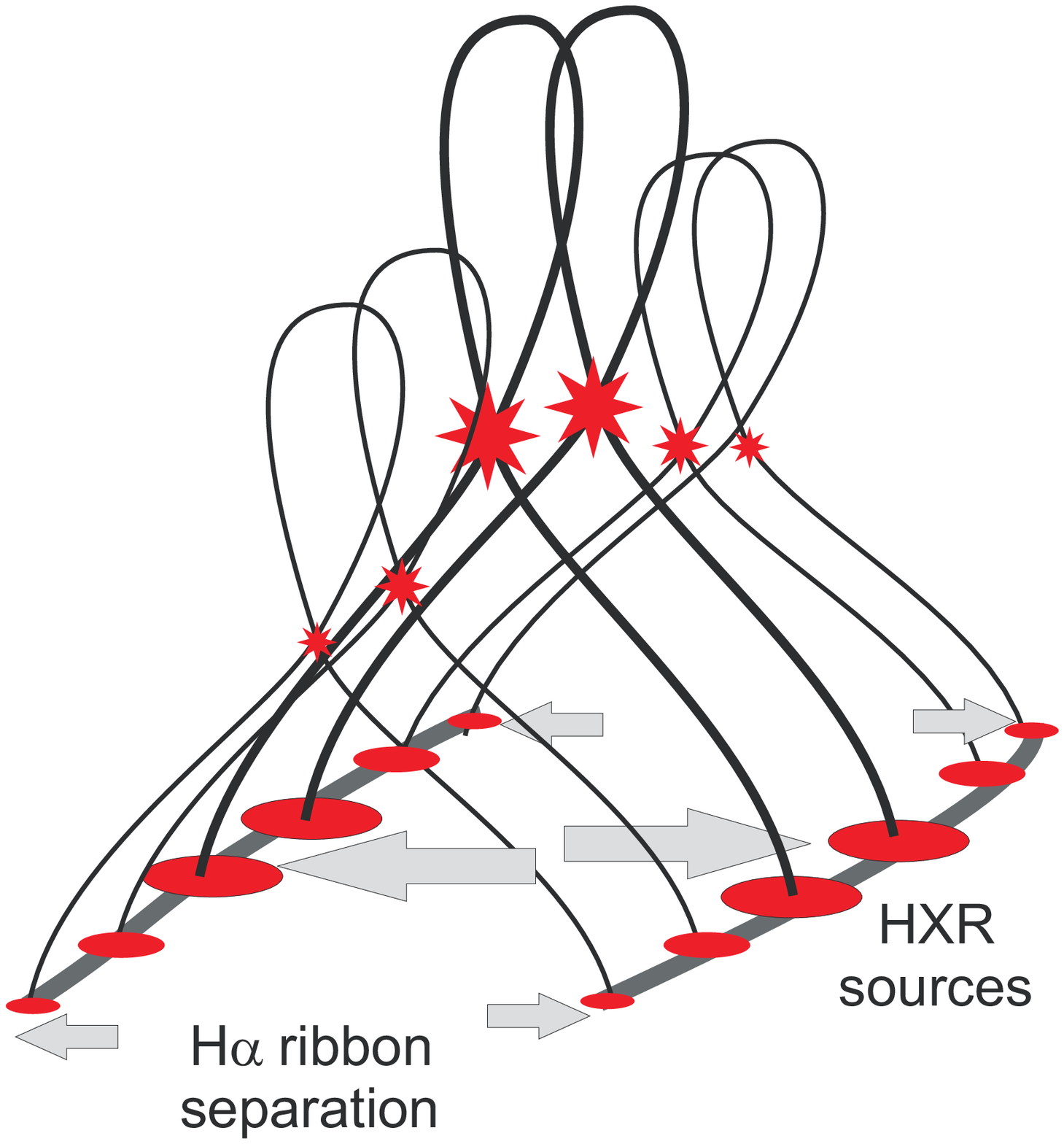,width=5cm}
\end{center}
\caption{\textit{Top}: Before reconnection - ``symmetric''
stretching of field lines with field lines in the middle get
stretched most. \textit{Bottom}: During reconnection - a higher
reconnection rate and stronger HXR sources are observed at the
center of the flaring arcade.}\label{arcade}
\end{figure}

Fig.~\ref{arcade} shows a scenario where due to a filament/CME
eruption (not drawn) a magnetic arcade rises up from the center
\citep[``symmetric eruption''; e.g.][]{tripathi06}\footnote{There
is also evidence for ``asymmetric eruptions'', e.g.
\cite{GrigisBenz05}} and the field lines get stretched mostly at
its center. After the reconnection process has started, electrons
are accelerated and produce HXR sources of different strength at
the footpoints of the loops along the arcade, namely stronger at
the center than at the flanks. We suppose that this might be due
to the associated current sheet (CS) that is larger at the center
part than at the flanks.

\begin{figure}
  \begin{center}
 \epsfig{file=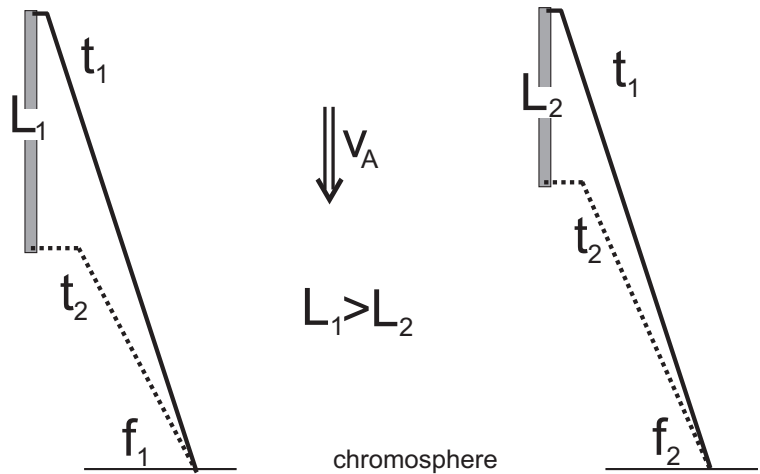,width=10cm}%
  \end{center}
  \caption{
Snapshots of single field lines that move during the reconnection
process through a CS (gray area) of length $L$. Field lines are
shown for two time steps ($t2$$>$$t1$ and $t2_{L1}$$>$$t2_{L2}$), namely right after
entering (solid lines) and before leaving the CS (dashed lines).
The outflow velocity is the same for both field lines (Alfv\'{e}n
velocity $v_{A}$).
  }\label{lee}
\end{figure}

In the following we will extract the effect of the CS length on
the energy released in the corona and subsequently deposited in
the chromosphere. When the magnetic field lines break and
reconnect, the magnetic energy is converted to heat, kinetic
energy and fast particle energy \citep[e.g.][]{forbes84}. We
assume an arcade of magnetic field lines embedded in a homogeneous
magnetic field in the chromosphere/photosphere and corona, that
goes through an ``idealized'' process of reconnection having a
similar reconnection rate along the entire arcade. From this
arcade we pick out two sets of neighboring field lines $f1$ and
$f2$, each mapping to a unit area in the chromosphere (cf.
Fig.~\ref{lee}). Each field line is swept into the reconnection
region and moves after reconnection (directed to the chromosphere)
along the CS of length $L$ with $v_{out}$$\approx$$v_{A}$
(external Alfv\'en speed). The total amount of energy $W$ supplied
to the chromosphere is proportional to the time $\Delta
t$$=$$L/v_{A}$ that the field line needs to move along the CS.
Assuming that the length of the CS through which $f1$ moves is
longer than for $f2$, i.e. $L_{1}$$>$$L_{2}$, the total amount of
energy supplied is higher for $f1$ than for $f2$ since $\Delta
t_{L1}$$>$$\Delta t_{L2}$. Another effect of the different CS
lengths at different parts of the arcade is that through the
larger area of the CS ($L_{1}$) a larger number of electrons can
be accelerated per unit time.

From the observations analyzed in Paper~I, we obtained that the
magnetic field strength is by a factor of 6--8 larger for those
parts of H$\alpha$ ribbons that were accompanied by HXR sources
compared to those without, whereas the propagation velocities of
different parts (with/without HXR sources) of the flaring ribbon
were found to be in the same range. Applying the above described
scenario, on the one hand a stronger magnetic field at the
footpoints reduces $\Delta t$ since it varies with $B^{-1}$
($v_{A}$$\propto$$B$ - minor effect) and on the other hand
increases the released energy ($W$$\propto$$B^{2}$ - major
effect). Moreover, the number of particles available for
acceleration is simply proportional to $L$.

\section{Discussion and Conclusion}

Concluding, the energy release as well as the number of
accelerated electrons varies with the length of the CS. Both favor
the occurrence of localized HXR footpoint sources. Thus, the
geometry of the CS might be an important parameter in ``guiding''
a large number of accelerated particles and enhanced energy
release to a subsystem of loops, which may be identified with the
confined HXR emission sources in contrast to the elongated ribbons
as observed in H$\alpha$ and UV. However, we also want to stress
that the magnetic field strength certainly plays a major role in
the occurrence of HXR source locations along the flaring arcade.

In several studies it was shown that a close correlation exists
between CME kinematics and associated flare characteristics
\citep{zhang01,vrsnak04,maricic06}. In the present study, a very
close match between the CME acceleration profile and the flare HXR
flux is seen. From this we propose that the CME might to a certain
degree influence the geometry of the CS, like the ``asymmetric''
or ``symmetric'' vertical stretching of the CS along the flaring
arcade.

To prove this proposal, it would be necessary to make a
quantitative study on flares where high cadence observations of
both the erupting filament and the corresponding HXR sources are
available. Flares that can be either associated to ``symmetric''
or ``asymmetric'' filament eruptions, should show us HXR source
locations at different parts along the ribbons \citep[see
also][]{GrigisBenz05, tripathi06}. This would provide further
evidence for the ``guiding/focusing'' of accelerated particles by
the CS geometry.

\section*{Acknowledgements}
M.T. gratefully acknowledges the Austrian {\em Fonds zur
F\"orderung der wissen\-schaftlichen Forschung} (FWF grant
J2512-N02). This work is supported by the Air Force Office of
Scientific Research, Air Force Material Command, USAF, under grant
number FA8655-06-1-3036.

\bibliographystyle{ceab}
\bibliography{zagreb2}

\end{document}